\documentclass[a4paper,12pt]{article}
\DeclareMathSizes{12}{12.5}{10}{10}
\usepackage[left=2.3cm,bottom=3cm,right=2.3cm,top=3cm]{geometry}
\usepackage{overpic,youngtab}
\usepackage{subfigure}
\usepackage[latin,english]{babel}
\usepackage{amsmath}
\usepackage{verbatim}
\usepackage{amssymb}
\usepackage{latexsym}
\usepackage{epsfig}
\usepackage{epstopdf}
\usepackage{scalerel}
\usepackage{graphics,psfrag,rotating}
\usepackage{graphicx}
\usepackage{dcolumn}
\usepackage{float}
\usepackage{pdflscape}
\usepackage{array}
\usepackage{booktabs}
\usepackage{amscd}
\usepackage{fancybox}
\usepackage{color}
\usepackage[colorlinks=true,citecolor=blue,,linktocpage=true,linkcolor=blue,urlcolor=black]{hyperref}
\usepackage{mathabx}

\definecolor{ogreen}{rgb}{0,0.7,0}

\DeclareGraphicsExtensions{.pdf,.png,.jpg,.gif,.jpeg}
\def\be{\begin{equation}}
\def\ee{\end{equation}}
\def\bea#1\eea{\begin{align}#1\end{align}}
\def\pd{\partial}
\def\a{\alpha}
\def\b{\beta}
\def\g{\gamma}
\def\d{\delta}
\def\e{\epsilon}

\def\m{\mu}
\def\n{\nu}
\def\t{\tau}

\def\l{\lambda}

\def\r{\rho}

\def\bR{\widebar{R}}

\def\bp{\widebar{\phi}}
\def\bn{\widebar{\nabla}}
\def\bR{\widebar{R}}

\def\s{\sigma}
\def\e{\epsilon}
\def\t{\tau}
\def\bi{\begin{itemize}}
	\def\ei{\end{itemize}}
\def\bg{\widebar{g}}
\def\bG{\widebar{\Gamma}}

\def\bph{\bar{\phi}}

\usepackage{authblk}
\usepackage{setspace}

\title{\centering\bf 
	One-Loop in first order quantum gravity}

\begin{document}


	\vspace*{-1cm}
	{\flushleft
		{{FTUAM-17-3}}
		\hfill{{IFT-UAM/CSIC-17-015}}}
	\vskip 1.5cm
	\begin{center}
		{\LARGE\bf One-loop divergences in first order Einstein-Hilbert gravity}\\[3mm]
		\vskip .3cm
		
	\end{center}
	\vskip 0.5  cm
	\begin{center}
		{\Large Enrique Alvarez}$^{~a}$, {\Large Jesus Anero}$^{~a}$
		{{\large and} \Large Raquel Santos-Garcia}$^{~a}$
		\\
		\vskip .7cm
		{
			$^{a}$Departamento de F\'isica Te\'orica and Instituto de F\'{\i}sica Te\'orica (IFT-UAM/CSIC),\\
			Universidad Aut\'onoma de Madrid, Cantoblanco, 28049, Madrid, Spain\\
			\vskip .1cm

			\vskip .5cm
			\begin{minipage}[l]{.9\textwidth}
				\begin{center} 
					\textit{E-mail:} 
					\tt{enrique.alvarez@uam.es},
					\tt{jesusanero@gmail.com},
					\tt{raquel.santosgarcia@estudiante.uam.es},
				\end{center}
			\end{minipage}
		}
	\end{center}
	\thispagestyle{empty}
	
	\begin{abstract}
		\noindent
One-loop counterterms are computed in the first order formalism for the Einstein-Hilbert action with a minimally coupled scalar field using the background field method and the heat kernel technique. The {\em off-shell}  divergent piece in the harmonic gauge is {\em exactly} the same as the one first found by 't Hooft and Veltman.
		
	\end{abstract}
	\newpage
	\tableofcontents

	\newpage
\section{Introduction}
There are many ways to write first order actions which are fully equivalent (at least classically) to the corresponding second order ones. They are usually more elaborate than the {\em naive} first order formalism, where the metric and the connection are treated as independent fields, insofar as they often need the introduction of auxiliary fields as Lagrange multipliers. This is a general mathematical result, valid for any system of ordinary differential equations.
\par
It is well-known that when considering the Palatini version of the Einstein-Hilbert action, that is, the {\em naive} first order formalism at the classical level, the connection is required to be the Levi-Civita one once the equations of motion are imposed. However, when more general quadratic in curvature metric-affine actions are considered in the first order approach, this relationship disappears even on-shell, hence allowing for more general connections. That is, the equations of motion do not force the connection to be the Levi-Civita one. This is of particular interest when analyzing quadratic theories in first order formalism, as these theories are quadratic in the derivatives of the connection and thus, no propagator decays faster than $\frac{1}{p^\text{{\tiny 2}}}$, indicating that there is still room for the theory to be unitary. Quadratic theories of gravity are renormalizable as opposed to General Relativity (cf. \cite{review_enrique} for a general review of quantum gravity and some approaches to the problem), so the study of this kind of theories in the first order formalism could give rise to a unitary and renormalizable theory of gravity \cite{quadratic_enrique,physicalcontent}. In this context, the computation to one-loop order for the linear Einstein-Hilbert action paves the way for future studies of more complex theories. 
\par
This work aims to revisit the computation of the one-loop quantum corrections to the gravitational action in the naive first order formalism. The action is assumed to be still the Einstein-Hilbert one, with the  Riemann tensor given solely in terms of the connection field and with the addition of a minimally coupled massless scalar field. We shall find exactly 't Hoot and Veltman's counterterm \cite{tHooft} even in the presence of a scalar field, and even off-shell, which is a result stronger than the one guaranteed by the general theorems of quantum field theory that only assert on-shell equality. The same problem was studied by Buchbinder and Shapiro \cite{Buchbinder} a while ago. More recently, first order formalism has been also studied for the Einstein-Hilbert action in \cite{mackeon,mikail}.
\par
The structure of the paper is as follows. In section 2, we expand the metric, the connection and the scalar field in their background value and a perturbation, obtaining the quadratic operators. After that, section 3 is devoted to the computation of the operator determining the counterterm and to the calculation of its heat kernel coefficient from which the counterterm is read. We also analyze in the same way the counterterm coming from the ghost action. Finally some conclusions are left for section 4. All the computations are made explicit so that the article can be self-contained and easy to follow.

	\section{The first order Einstein-Hilbert action with a scalar field}
We consider the Einstein-Hilbert action with the metric coupled to a massless scalar field $\phi$ given by
\bea
S_{\text{\tiny{EH$\phi$}}}&\equiv -\frac{1}{2}\int
d^n x\sqrt{|g|}g^{\m\n}R_{\m\n}+\frac{1}{2}\int
d^n x\sqrt{|g|}\frac{1}{2}g^{\m\n}\partial_\m\phi\partial_\n\phi,
\eea
where we take the gravitational coupling constant as unity, namely, $\kappa=1$.
The metric, the connection  and the scalar field are treated as independent fields and  are expanded in a background field and a perturbation as
\bea &g_{\m\n}=\bg_{\m\n}+ h_{\m\n},\nonumber\\
&\Gamma^\l_{\m\n}=\bG^\l_{\m\n}+ A^\l_{\m\n},\nonumber\\
&\phi=\bph+\phi.
\eea
Let us note that indices are raised with the background metric, and the quantities computed with respect to this metric have a bar. We also take as the background connection the Levi-Civita connection built from the background metric.\footnote{In this sense, the equation of motion for the connection field is already imposed from the beginning of the computation.}
As usual, linear terms cancel provided the classical fields obey the  background equations of motion (EoM) given by
\bea \label{EM}&\frac{1}{2}\bg_{\m\n}\bR-\bR_{\m\n}-\frac{1}{4}\bg_{\m\n}\partial_\l\bph\partial^\l\bph+\frac{1}{2}\partial_\m\bph\partial_\n\bph=0,\nonumber\\ 
&\bg^{\a\b}\bar{\nabla}_\lambda \bg_{\a\b}=0, \nonumber \\
&\bar{\Box}\bph=0.
\eea
To take into account the one-loop effects it is enough to expand the action up to quadratic order in the perturbations so that this piece reads
\bea 
\label{BF}S_{\text{\tiny{2}}}&=\int
d^n x~\sqrt{|\bg|}~\left\{h^{\a\b} M_{\a\b\g\e} h^{\g\e}+h^{\a\b}\vec{N}_{\a\b~\t}^{~~~\g\e}A^{\t}_{\g\e}-A^{\lambda}_{\alpha\beta}\vec{N}_{\l~\g\e}^{\a\b~~}h^{\g\e}+\right.\nonumber\\
&\left.+A^{\l}_{\a\b}K^{\a\b~\g\e}_{~\l~~\t}A^{\t}_{\g\e}+h^{\a\b}E_{\a\b}\phi+\phi F \phi\right\}.
\eea
The symbol $\vec{N}$ is used to indicate the fact that the derivative contained in the operator acts on the right. The explicit expression for these operators is then
\bea\label{K}
M_{\a\b\g\e}&= \frac{1}{16}\left(\bg_{\a\e}\bg_{\b\g}+\bg_{\a\g}\bg_{\b\e}-\bg_{\a\b}\bg_{\g\e}\right)\left(\bR-\frac{1}{2}\bg^{\r\s}\pd_\r\bph\pd_\s\bph\right)
+\nonumber\\
&+\frac{1}{8}\left(\bg_{\a\b}\bR_{\g\e}+\bg_{\g\e}\bR_{\a\b}-\bg_{\a\g}\bR_{\b\e}-\bg_{\a\e}\bR_{\b\g}-\bg_{\b\g}\bR_{\a\e}-\bg_{\b\e}\bR_{\a\g}\right)-\nonumber\\
&-\frac{1}{16}\left(\bg_{\a\b}\pd_\g\bph\pd_\e\bph+\bg_{\g\e}\pd_\a\bph\pd_\b\bph-\bg_{\a\g}\pd_\b\bph\pd_\e\bph-\bg_{\a\e}\pd_\b\bph\pd_\g\bph-\bg_{\b\g}\pd_\a\bph\pd_\e\bph-\bg_{\b\e}\pd_\a\bph\pd_\g\bph\right)\nonumber\\
N_{\g\e~\l}^{~~\a\b}&=\frac{1}{8}\left(\d^\a_\g\delta^{\beta}_{\e}+\d^\a_\e\delta^{\beta}_{\g}-\bg_{\g\e}\bg^{\a\b}\right)\bar{\nabla}_\lambda
+\nonumber\\
&+\frac{1}{16}\left(\bg_{\g\e}\d^\b_\l\bar{\nabla}^\a -\d^\a_\g\delta^{\beta}_{\lambda}\bar{\nabla}_\e-\d^\a_\e\delta^{\beta}_{\lambda}\bar{\nabla}_\g
+\bg_{\g\e}\d^\a_\l\bar{\nabla}^\b -\d^\b_\g\delta^{\a}_{\lambda}\bar{\nabla}_\e-\d^\b_\e\delta^{\a}_{\lambda}\bar{\nabla}_\g\right) \nonumber\\
K^{\g\e~\a\b}_{~\t~~\l}&= \frac{1}{8}\Big[
\d^\b_\t \d^\g_\l \bg^{\a\e}
+\d^\b_\t \d^\e_\l \bg^{\a\g}
+\d^\a_\t \d^\e_\l \bg^{\b\g}
+\d^\a_\t \d^\g_\l \bg^{\b\e}-\d^\e_\t \d^\g_\l \bg^{\a\b}-\d^\g_\t \d^\e_\l \bg^{\a\b}-\d^\b_\l \d^\a_\t \bg^{\g\e} -\d^\a_\l \d^\b_\t \bg^{\g\e}\Big]
\nonumber\\
E_{\a\b}=&\frac{1}{4}\bg_{\a\b}\bg^{\r\s}\pd_\r\bph\pd_\s-\frac{1}{4}\pd_\a\bph\pd_\b-\frac{1}{4}\pd_\b\bph\pd_\a\nonumber\\
F=-&\frac{1}{4}\bar{\Box}\nonumber\\
\eea

\section{Computation of the counterterms}
In order to compute the counterterm we need to take the effective action as the starting point, which in this case depends on the three fields appearing in the theory
\bea
e^{iW \scriptstyle\left[\bg_{\m\n},\bG^\l_{\r\s}, \bph\scriptstyle\right]}&=\int \mathcal{D}h\mathcal{D}A \mathcal{D} \phi~e^{iS_{\text{\tiny 2}}[h,A,\phi]}\eea
Taking advantage of the form of the background expansion (\ref{BF}), we can rewrite the metric and connection pieces by completing squares as
\be hMh+h\vec{N}A-A\vec{N}h+AKA
=hMh+[h\vec{N}+AK]K^{-1}[-\vec{N}h+KA]+h\vec{N}K^{-1}\vec{N}h.
\label{integrationbyparts}
\ee 
Due to the translational invariance of the integration measure, we can redefine the connection field perturbations so that the second term in \eqref{integrationbyparts} becomes quadratic in those. The integral over the quantum connection fields, ${\cal D}A$, is then a trivial gaussian integral yielding
\bea
e^{iW}=\int \mathcal{D}h \mathcal{D}\phi~e^{\left\{i\int
	d^n x~\sqrt{|g|}~h^{\a\b}\left(M_{\a\b\g\e}+D_{\a\b\g\e}\right)h^{\g\e}+h^{\a\b}E_{\a\b}\phi+\phi F \phi
	\right\}},\label{PI}\eea
where we have defined the new piece of the quadratic operator mediating between the metric perturbations as $D_{\a\b\g\e}\equiv \vec{N}_{\a\b~\l}^{~~~\m\n}(K^{-1})_{\m\n~\r\s}^{~\l~~\t}\vec{N}_{~\t~\g\e}^{\r\s}$.
\par
This integration before fixing the gauge is only possible if we are able to invert the operator $K$. This is indeed possible and the explicit expression for $K^{-1}$ yields
\bea
(K^{-1})_{\a\b~\g\e}^{~\l~~\t}&=-\dfrac{2}{n-2}\left\{ \delta_{\gamma}{}^{\tau} \delta_{\epsilon}{}^{\lambda} \bg_{\alpha \beta} +  \delta_{\gamma}{}^{\lambda} \delta_{\epsilon}{}^{\tau} \bg_{\alpha \beta} + \delta_{\alpha}{}^{\tau} \delta_{\beta}{}^{\lambda} \bg_{\gamma \epsilon} +  \delta_{\alpha}{}^{\lambda} \delta_{\beta}{}^{\tau} \bg_{\gamma \epsilon} \right\}+ \nonumber\\
& +  \delta_{\beta}{}^{\tau} \delta_{\epsilon}{}^{\lambda} \bg_{\alpha \gamma} + \delta_{\alpha}{}^{\tau} \delta_{\epsilon}{}^{\lambda} \bg_{\beta \gamma} + \delta_{\beta}{}^{\tau} \delta_{\gamma}{}^{\lambda} \bg_{\alpha \epsilon}+ \delta_{\alpha}{}^{\tau} \delta_{\gamma}{}^{\lambda} \bg_{\beta \epsilon} +  \nonumber\\&+  \dfrac{2}{n^2-3n+2} \left\{\delta_{\beta}{}^{\lambda} \delta_{\epsilon}{}^{\tau} \bg_{\alpha \gamma} +  \delta_{\alpha}{}^{\lambda} \delta_{\epsilon}{}^{\tau} \bg_{\beta \gamma}+   \delta_{\beta}{}^{\lambda} \delta_{\gamma}{}^{\tau} \bg_{\alpha \epsilon} + \delta_{\alpha}{}^{\lambda} \delta_{\gamma}{}^{\tau} \bg_{\beta \epsilon}\right\} - \nonumber\\&-\bg_{\alpha \epsilon} \bg_{\beta \gamma} \bg^{\lambda \tau} - \bg_{\alpha \gamma} \bg_{\beta \epsilon} \bg^{\lambda \tau}+  \dfrac{2}{n-2} \bg_{\alpha \beta} \bg_{\gamma \epsilon} \bg^{\lambda \tau},
\eea
in such a way that
\bea
D_{\a\b\g\e}&=\frac{1}{16}(2\bg_{\a\b}\bg_{\g\e}-\bg_{\a\g}\bg_{\b\e}-\bg_{\a\e}\bg_{\b\g})\bar{\Box}-\frac{1}{8}(\bg_{\a\b}\bn_\g \bn_\e+\bg_{\g\e}\bn_\a \bn_\b)+\nonumber\\
&+\frac{1}{16}(\bg_{\b\e}\bn_\g \bn_\a+\bg_{\a\e}\bn_\g \bn_\b+\bg_{\a\g}\bn_\e \bn_\b+\bg_{\b\g}\bn_\e \bn_\a).
\eea
Before continuing, one may ask why it is the case that we can invert the operator $K$ before gauge fixing, that is, not having to take care of any zero modes. We discuss this point in Appendix (\ref{C}), where we have done an exhaustive study of the gauge symmetry of the whole action.
\par
Nonetheless, the quadratic action that we have left is still invariant under the quantum gauge symmetry corresponding to diffeomorphism invariance given by the transformations
\bea
\d h_{\m\n}&=\bn_{\m}\xi_{\n}+\bn_{\n}\xi_{\m}+\mathcal{L}_\xi h_{\m\n}\nonumber\\
\d \phi &= \xi^\mu \bn_\mu \phi 
\eea
Let us  consequently define the de Donder or harmonic gauge fixing 
\be
S_{\text{\tiny{gf}}}=\frac{1}{4}\,\,\int\,d^nx\,\,\sqrt{\bg}\,\bg_{\m\n}\chi^\m\chi^\n,
\ee
where 
\be
\chi_\n=\bn^\m h_{\m\n}-\frac{1}{2}\bn_\n h-\phi\pd_\n\bp.
\ee	
The quadratic action after adding the gauge part then reads
\bea
S_{\text{\tiny{2+gf}}}=\frac{1}{4}\,\int\,d^nx\,\sqrt{\bg}\,\psi^A\Delta_{AB}\psi^B,
\eea 
where we have written the quadratic operator corresponding to the generalized field, $\psi^A$, defined as the vector
\be
\psi^A\equiv\begin{pmatrix}
	h^{\m\n}\\\phi
\end{pmatrix},
\ee
and the operator takes the form
\be
\Delta_{AB}=-g_{AB}\bar{\Box}+Y_{AB}.
\label{op}
\ee
Let us specify its  different pieces. The term multiplying the box operator acts as a sort of internal metric $g_{AB}$ and reads
\be
g_{AB}=\begin{pmatrix}
	C_{\a\b\m\n}&0\\0&1 \end{pmatrix},
\ee
with $C_{\m\n\r\s}=\frac{1}{4}(\bg_{\m\r}\bg_{\n\s}+\bg_{\m\s}\bg_{\n\r}-\bg_{\m\n}\bg_{\r\s})$. The components of $Y_{AB}$ are also detailed below for completeness
\bea
&Y^{hh}_{AB}=\frac{1}{4}(\bg_{\m\r}\bg_{\n\s}+\bg_{\m\s}\bg_{\n\r}-\bg_{\m\n}\bg_{\r\s})\left(\bR-\frac{1}{2}\bg^{\a\b}\pd_\a\bph\pd_\b\bph\right)-\frac{1}{2}\left(\bR_{\m\r\n\s}+\bR_{\n\r\m\s}\right)+\nonumber\\
&+\frac{1}{2}\left(\bg_{\m\n}\bR_{\r\s}+\bg_{\r\s}\bR_{\m\n}\right)-\frac{1}{4}\left(\bg_{\m\r}\bR_{\n\s}+\bg_{\m\s}\bR_{\n\r}+\bg_{\n\r}\bR_{\m\s}+\bg_{\n\s}\bR_{\m\r}\right)\nonumber\\
&-\frac{1}{4}\left(\bg_{\m\n}\pd_\r\bph\pd_\s\bph+\bg_{\r\s}\pd_\m\bph\pd_\n\bph-\bg_{\m\r}\pd_\n\bph\pd_\s\bph-\bg_{\m\s}\pd_\n\bph\pd_\r\bph-\bg_{\n\r}\pd_\m\bph\pd_\s\bph-\bg_{\n\s}\pd_\m\bph\pd_\r\bph\right),\nonumber\\
&Y^{h\phi}_{AB}=Y^{\phi h}_{AB}=\frac{1}{2}\left[\partial_\a\partial_\b\bph+\partial_\b\partial_\a\bph-\bg_{\a\b}\bar{\Box}\bph\right],\nonumber\\
&Y^{\phi\phi}_{AB}=\bg^{\r\s}\partial_\r\bph\partial_\s\bph.\eea
The form of the usual four-dimensional one-loop counterterm is tabulated in  \cite{Barvinsky} for minimal operators of the same type as \eqref{op}, namely,
\bea\label{coe}
\Delta S&=\frac{1}{(4\pi)^2}\frac{1}{\e}\int d^n x~\sqrt{|\bg|}\text{tr}\Big\{\frac{1}{360}\left(2\bR_{\m\n\r\s}\bR^{\m\n\r\s}-2\bR_{\m\n}\bR^{\m\n}+5\bR^2\right) \mathbb{I}+\nonumber\\
&+\frac{1}{2}Y^2-\frac{1}{6}\bR Y+\frac{1}{12}W_{\m\n}W^{\m\n}\Big\},\eea
where the field strength is defined through
\be
[\bn_\m,\bn_\n]h^{\a\b}=W_{~~~\r\s\m\n}^{\a\b}h^{\r\s},
\ee
with
\be W_{~~~\r\s\m\n}^{\a\b}=\frac{1}{2}\left(\d^\b_\r\bR^\a_{~\s\m\n}+\d^\b_\s\bR^\a_{~\r\m\n}+\d^\a_\r\bR^\b_{~\s\m\n}+\d^\a_\s\bR^\b_{~\r\m\n}\right),\ee
which is symmetric under ${\a\b \leftrightarrow \r\s}$. Let us note that the trace in \eqref{coe} also encodes the trace of the matrix $\Delta_{AB}$. With this, only a few traces need to be computed in order to find the explicit value of the counterterm. These are given by
\bea
&\text{tr}~ \mathbb{I} =\frac{n(n+1)}{2}+1\nonumber\\
&\text{tr}~Y=g^{AB}Y_{AB}
=\frac{n(n-1)}{2}\bR+\frac{8+3n-n^2}{4}\bg^{\r\s}\partial_\r\bph\partial_\s\bph\nonumber\\
&\text{tr}~Y^2=Y_{AB}\,g^{BC}\,Y_{CD}\,g^{DA}=3\bR_{\m\n\r\s}\bR^{\m\n\r\s}+\frac{n^2-8n+4}{n-2}\bR_{\m\n}\bR^{\m\n}+\frac{n^3-5n^2+8n+4}{2(n-2)}\bR^2-\nonumber\\
&\quad\quad\quad -\left[\frac{2n(n-4)}{ (n-2)}+4\right]\bR^{\m\n}\partial_\m\bph\partial_\n\bp+\frac{n^3-7n^2+10n+8}{2(2-n)}\bR\bg^{\r\s}\partial_\r\bph\partial_\s\bph +2\bar{\Box}\bph\bar{\Box}\bph\nonumber\\
&\quad\quad\quad+\frac{n^3-n^2+14n-40}{8(n-2)}\left(\bg^{\r\s}\pd_\r\bp\pd_\s\bp\right)^2\nonumber\\
&\text{tr}~W_{\m\n}W^{\m\n}=-(n+2)\bR_{\m\n\r\s}\bR^{\m\n\r\s}
\eea
Using expression (\ref{coe}), the full gauge, gravitational and scalar field contributions to the one-loop counterterm are given by
\bea\label{a2}
\Delta S_{\text{\tiny{2+gf}}}&=\frac{1}{(4\pi)^2}\frac{1}{\e}\int d^n x~\sqrt{|\bg|}\frac{1}{360}\Bigg\{\left(482-29n+n^2\right)\bR_{\m\n\r\s}\bR^{\m\n\r\s}+\nonumber\\
&+\frac{724-1440n+181n^2-n^3}{n-2}\bR_{\m\n}\bR^{\m\n}+
\frac{5(140+264n-145n^2+25n^3)}{2(n-2)}\bR^2-\nonumber\\
&-\frac{360(-4-2n+n^2)}{ n-2}\bR^{\m\n}\partial_\m\bph\partial_\n\bph-\frac{15(32+62n-37n^2+5n^3)}{n-2}\bR\bg^{\r\s}\partial_\r\bph\partial_\s\bph+\nonumber\\
&+\frac{45(n^3-n^2+14n-40)}{2(n-2)}\left(\bg^{\r\s}\partial_\r\bph\partial_\s\bph\right)^2+360\bar{\Box}\bph\bar{\Box}\bph\Bigg\}.
\eea
\par
Finally, the contribution coming from the ghost loops is also needed. The quadratic ghost lagrangian then reads
\be
S_{\text{\tiny{gh}}}=\frac{1}{4}\,\,\int\,d^nx\,\,\sqrt{\bg}\,V^*_\m\left(-\bg^{\m\n}\bar{\Box}-\bR^{\m\n}+\bn^\m\bph\bn^\n\bph\right)V_\n,
\ee
whre operators cubic in the fluctuations are not taken into account as they are irrelevant at one-loop (the ghosts being quantum fields do not appear as external states). The corresponding Laplacian operator is simply given by
\be \Delta_{\m\n}=-\bg_{\m\n}\bar{\Box}+Y_{\m\n},\ee 
with
\be Y_{\m\n}=-\bR_{\m\n}+\bn_\m\bph\bn_\n\bph.\ee
We can compute the traces in the same way as before so that
the complete ghost contribution can be then read from (\ref{coe}) yielding
\bea\label{a2g}
\Delta S_{\text{\tiny{gh}}}&=\frac{1}{(4\pi)^2}\frac{1}{\e}\int d^n x~\sqrt{|\bg|}\frac{1}{360}\left[(2n-30)\bR_{\m\n\r\s}\bR^{\m\n\r\s}+(180-2n)\bR_{\m\n}\bR^{\m\n}+(5n+60)\bR^2-\right.\nonumber\\
&\left.-360 \bR^{\m\n}\partial_\m\bph\partial_\n\bph-60\bR \bg^{\r\s}\partial_\r\bph\partial_\s\bph+180\left(\bg^{\r\s}\partial_\r\bph\partial_\s\bph\right)^2\right].
\eea
Adding the two pieces (\ref{a2}) and (\ref{a2g}) (note the factor and the sign of the ghost contribution), and specifying the result to $n=4$, the full one-loop counterterm reads 
\bea\label{countfull}
\Delta S_{\text{\tiny{EH$\phi$}}}&=\frac{1}{(4\pi)^2}\frac{1}{\e}\int d^n x~\sqrt{|\bg|}\left(\text{tr}\, a_2 \left(x,x\right)-2\text{tr}\, a^{\text{gh}}_2 \left(x,x\right)\right)=\nonumber\\&=\frac{1}{(4\pi)^2}\frac{1}{\e}\int d^n x~\sqrt{|\bg|}\left\{\frac{43}{60}\bR_{\m\n}\bR^{\m\n}+\frac{1}{40}\bR^2+\frac{1}{6}\bR \bg^{\r\s}\pd_\r\bph\pd_\s\bph+\left(\bg^{\r\s}\pd_\r\bp\pd_\s\bp\right)^2+\bar{\Box}\bph\bar{\Box}\bph\right\},
\eea
where the well-known four-dimensional Gauss-Bonnet identity has been used, namely,
\be 
\label{GB}\bR_{\m\n\r\s}\bR^{\m\n\r\s}-4\bR_{\m\n}\bR^{\m\n}+\bR^2=\text{total derivative}.
\ee
Substituting the background equations of motion, the {\em on-shell} effective action is finally obtained
\bea
\label{count}\Delta S_{\text{\tiny{EH$\phi$}}}&=\frac{1}{(4\pi)^2}\frac{1}{\e}\int d^n x~\sqrt{|\bg|}\frac{203}{40}\bR^2,
\eea
which exactly matches 't Hooft and Veltman's result\cite{tHooft}. This result is at variance, however, with the one obtained in \cite{Buchbinder}.
\section{Conclusions}
We have analyzed the Einstein-Hilbert action with a massless scalar field in the {\em naive first order formalism}. This action is the same as the second order action, the only change being that the metric and the connection are now treated as independent fields. We have computed the one-loop divergences, using standard background field and heat kernel techniques. We find that the one-loop corrections for the Einstein-Hilbert action are identical in the first order and second order formalisms when using the same field parametrizations and the same gauge fixing, as first computed in the classic paper by 't Hooft and Veltman. This problem was already studied in \cite{Buchbinder}, but our results slightly differ from this reference.
\par
Although it is well known that at the classical level first order and second order formalisms coincide, we have shown that this equivalence also holds at the one-loop level. Moreover, we find that this is true even off-shell, that is, before imposing the background EoM. This is more than we had the right to expect as the general theorems only ensure that the effective actions should coincide on-shell, this being enough to ensure equality of the S-matrices. Here, however, the EoM are identical in the first and second order approaches, so that the on-shell equality implies the off-shell one as well, as long as we choose the same gauge fixing and field parametrizations.

\section{Acknowledgements}
This work has been partially supported by the Spanish Research Agency (Agencia Estatal de Investigacion) through the PID2019-108892RB-I00/AEI/ 10.13039/501100011033 grant as well as the IFT Centro de Excelencia Severo Ochoa SEV-2016-0597 one, and the European Union's Horizon 2020 research and innovation programme under the Marie Sklodowska-Curie grants agreement No 674896 and No 690575. RSG is supported by the Spanish FPU Grant No FPU16/01595.
\newpage
\appendix

\newpage
 
\end{document}